\documentclass[prd, amsfonts, twocolumn, nofootinbib, showpacs]{revtex4}
\usepackage{graphicx}
\usepackage{epsfig}
\usepackage{color}
\usepackage{amsmath}
\usepackage{amssymb}
\usepackage{physics}
\newcommand{\be}{\begin{equation}}
\newcommand{\ee}{\end{equation}}
\newcommand{\bea}{\begin{eqnarray}}
\newcommand{\eea}{\end{eqnarray}}

\newcommand{\gapp}{\mathrel{\raise.3ex\hbox{$>$}\mkern-14mu
\lower0.6ex\hbox{$\sim$}}}
\newcommand{\lapp}{\mathrel{\raise.3ex\hbox{$<$}\mkern-14mu
\lower0.6ex\hbox{$\sim$}}}
\def\bbox{{\,\lower0.9pt\vbox{\hrule \hbox{\vrule height 0.2 cm
\hskip 0.2 cm \vrule  height 0.2 cm}\hrule}\,}}

\begin{document}
\title{State of a particle pair produced by the Schwinger effect is not necessarily a maximally entangled Bell state}
\author{De-Chang Dai\footnote{corresponding author: De-Chang Dai,\\ email: diedachung@gmail.com\label{fnlabel}}, }
\affiliation{ Center for Gravity and Cosmology, School of Physics Science and Technology, Yangzhou University, 180 Siwangting Road, Yangzhou City, Jiangsu Province, P.R. China 225002 }
\affiliation{  Department of Physics, Case Western Reserve University,
10900 Euclid Avenue, Cleveland, OH 44106 }


\begin{abstract}
\widetext

We analyze the spins of a Schwinger particle pair in a spatially uniform but time dependent electric field. The particle pair's spins are in the maximally entangled Bell state only if the particles' momenta are parallel to the electric field. However if transverse momentum is present, the spins are not in the maximally entangled Bell state.
 The reason is that the pair is created by the external field, which also carries angular momentum, and the particle pair can take away some of this external angular momentum. 
\end{abstract}

\pacs{}
\maketitle

\section{ Introduction.}
The action of an electron in a constant electric field had been formulated by Sauter, Heisenberg and Euler\cite{1931ZPhy...69..742S,1936ZPhy...98..714H}. Schwinger used this action in a gauge invariant form in a constant electromagnetic field,
and found the effect of charged particle pair production if the electric field is higher than the critical value $E_{cri}=\frac{m_e^2c^3}{q_e\hbar}$\cite{1951PhRv...82..664S,1949PhRv...76..749F}.  Recently, in experiments with  high power lasers, this critical value was achieved, so experimental study of the Schwinger effect might be realized in a foreseeable future. Apart from a constant electric field background, additional efforts have been focused on  a spatially dependent strong fields\cite{2010PhRvD..82b5015K,2017PhRvD..96e6017L}, time dependent fields\cite{BialynickiBirula:2011eg}, thermal backgrounds\cite{2012PhRvD..86l5007K}, multipair creation states\cite{2016PhLB..760..552W,2017PhRvD..96e6017L}, strong electric and magnetic fields\cite{2008PhRvD..78j5013K}, and particle creation in pulsars\cite{2012MNRAS.420.1673L}.

A Schwinger pair is a pair of virtual particles separated by an external field to become a real pair. Since even virtual particle should conserve quantum numbers, they are assumed to be highly correlated or entangled. Especially, the pair's spin state is generally expected to be one of the maximally entangled Bell states\cite{Ebadi:2014ufa, Li:2016zyv,Mathur:2009hf}
\begin{equation}
\label{entangle}
\frac{1}{\sqrt{2}}(\ket{\uparrow\downarrow}\pm \ket{\downarrow\uparrow}) .
\end{equation}
Here $\uparrow$ and $\downarrow$ represent the particles' spin directions.  We can consider the first particle to be an electron and the second one positron. If the particles' states are represented by eq. \eqref{entangle}, then if one of the particles' states is known, the other particle's state is also known. This correlation is one of the reason why the Bell state is called a maximally entangled state. Even though it is intuitive to assume that a Schwinger pair's spin state is in one of the Bell states, the pair production is derived in a basis different from the spin basis\cite{Ebadi:2014ufa, Li:2016zyv}. We will see that the Schwinger pair is not in a Bell state after transforming it to the correct spin basis.

Apart from strong electromagnetic field, a strong gravitational field can also be a source of particle pairs, in a process similar to the Schwinger effect.  Davies et al. calculated the energy momentum tensor near a black hole\cite{Davies:1976ei}. They found that there is negative energy flux into the black hole, and positive energy flux to infinity, which is Hawking radiation. This is an explicit evidence that Hawking radiation is created by gravitational particle pair production, which is just a variant of the Schwinger effect. If a particle pair is separated by gravity, one member of the pair may fall into the horizon and the other may run away from the horizon. The runaway particle eventually becomes part of Hawking radiation at infinity. Since a Schwinger particle pair is assumed to be highly entangled, the pair produced in the Hawking effect is also assumed to be highly entangled. This assumption, combined with an assumption that gravity is a local theory, led to the so-called information loss paradox\cite{Mathur:2009hf}. In addition, one of the assumptions of the Firewall paradox in black hole physics is tightly connected to particle entanglement\cite{Almheiri:2012rt}. However, it has been noted that the entanglement is observer dependent \cite{FuentesSchuller:2004xp,Alsing:2006cj,Adesso:2007wi},  and the pair can be disentangled after propagating some distance\cite{MartinMartinez:2010ar}.

Although it is natural to assume that a virtual particle pair is highly entangled, one cannot avoid interaction with an external field which makes them real. This external field can carry momentum, angular momentum and other quantum numbers. Therefore one cannot consider only an isolated virtual particles pair. The particle pair can carry also its own angular momentum, apart from the individual spins of particles. The actual spin state of the pair can be different from the expected virtual pair state in eq (\ref{entangle}), and can actually be
\begin{equation}
\label{portion}
A_0\ket{\uparrow\downarrow}+A_1\ket{\downarrow\uparrow}+A_2\ket{\uparrow\uparrow}+A_3\ket{\downarrow\downarrow} .
\end{equation}
 We again consider the first particle to be an electron and the second one positron. If this description is true, one cannot know the spin of one particle based on the knowledge of the spin of the other particle without an actual direct detection.

\begin{figure}
\includegraphics[width=8cm]{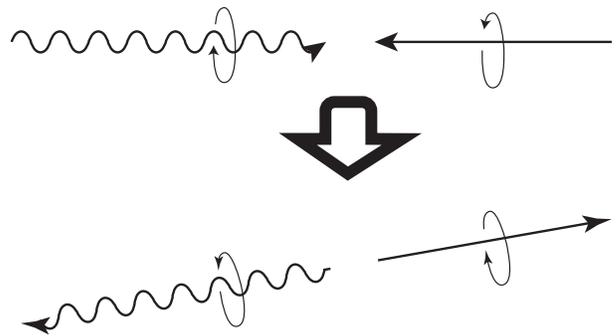}
\caption{Compton scattering in the center of mass frame. A right handed photon collides with a right handed electron. They turn almost completely backward. Their helicities are still both right handed. The total spin is not conserved, because the reflected waves are the p-waves, which take away angular momentum.
}
\label{scatter}
\end{figure}

 We note that an apparent difference between the total initial and final spin in some process is very common. For example, consider a head-on electron-photon collision (fig \ref{scatter}). A right handed photon and a right handed electron collide and turn almost completely backward. The reflected photon and electron are both right handed. The total spin appears not to be conserved  because there is angular momentum involved in the process. 

The Swinger pair's spin creation rate can be studied by using the Dirac-Heisenberg-Wigner (DHW) function\cite{1991PhRvD..44.1825B,2010PhRvD..82j5026H,2011PhRvD..83f5020B,BialynickiBirula:2011eg}. However, since we want to know a single pair's spin correlation,  Bogoliubov transform can give us a better description than the DHW method. The ``in" vacuum is based on the equation of motion in a constant electric field background. The ``out" vacuum is based on the particle's spin up and down states. We will now show that if the particles' momenta are parallel to the electric field, the pair's spin is in a Bell-like state, and one can know the exact complete particle state just by knowing the quantum state of one particle. However, if the particle pair has transverse momentum (with respect to the external field), then pair's spins are not in a Bell-like state. One cannot know one particle state solely based on the information from the other particle. This implies the external field does affect the Schwinger pair's quantum state. This also implies that Hawking radiation should not be treated as a local phenomenon (see also discussion around eq. 4 in \cite{Hutchinson:2013kka}). In the foreseeable future, the high intensity lasers may produce electric fields above the Schwinger pair production threshold. Recent studies mainly focus on the pair production numbers\cite{BialynickiBirula:2011eg}. However, the laser's photons carry spins, therefore the electron and positron pair's spins will also depend on the annihilated photons states. This is the effect that we want to examine here. In principle, we expect that the particle pair's spin state depends on the scattering angle. In the following we review the quantization procedure and ``out" vacuum . We then calculate the amplitude of different particle pairs' spin states and helicities.

\section{ Quantization }
Schwinger particle pairs are created by the strong electric field. Both charged fermions and bosons can be created in the process. Since we want to study whether the particles' spins are highly entangled, we focus on fermion pairs. In particular, we study electron-positron pairs.

Here we follow Klunger et al.'s study in \cite{Kluger:1992gb}. The Lagrangian density for electrodynamics is
\begin{equation}
L=\bar{\psi} i \gamma^\mu (\partial_\mu +ieA_\mu)\psi-m\bar{\psi} \psi-\frac{1}{4}F_{\mu\nu}F^{\mu\nu}
\end{equation}
where the metric convention is taken to be $(+,-,-,-)$. $\psi$ is a charged Dirac field, while $A^\mu$ is the background electromagnetic field. The $\gamma$ matrices are
\begin{equation}
\gamma^0 = \begin{bmatrix}
       I &  0           \\[0.3em]
       0           & -I
     \end{bmatrix}\text{, } \gamma^i = \begin{bmatrix}
       0 &  \sigma^i           \\[0.3em]
       -\sigma^i           & 0
     \end{bmatrix}
\end{equation}
where, $i=1,2,3$. $I$ is a 2 by 2 unit matrix. $\sigma^i$ are the Pauli matrices. The equation of motion for $\psi$ is
\begin{eqnarray}
\label{eqm}
(i\gamma^\mu \partial_\mu -e \gamma^\mu A_\mu -m)\psi =0 .
\end{eqnarray}
$\psi$ can be expressed through a new field $\phi$ as
\begin{equation}
\label{transform}
\psi =(i\gamma^\mu \partial_\mu -e\gamma^\mu A_\mu +m )\phi.
\end{equation}

 Eq. (\ref{eqm}) becomes a quadratic Dirac equation
\begin{equation}
\Big[ (i\partial_\mu-eA_\mu)^2 -\frac{e}{2}\sigma^{\mu\nu}F_{\mu\nu}-m^2\Big]\phi=0 .
 \end{equation}

In general, $A^\mu$ is space and time dependent. However, to simplify the discussion, we consider spatially uniform electric field which points to the $z$ direction.  The none-zero $A^\mu$ component is $A^3=a(t)$. Before the electric field is applied, the electric field is $0$, and $a (-\infty)=0$. After the electric field is turned off, $A^\mu$ becomes constant, so $\lim_{t\rightarrow \infty}a(t)=\text{constant}$. Then the equation can be simplified to

\begin{equation}
\Big[ \partial_\mu\partial^\mu +e^2 a^2+2ia\partial_3 -ie\partial_0 a \gamma^0\gamma^3+m^2\Big]\phi=0 .
\end{equation}

Spatial homogeneity implies that the solutions can be written in the form of
\begin{equation}
\phi_{\boldsymbol{k},j}=e^{i \boldsymbol{k}\cdot \boldsymbol{x}}f_{\boldsymbol{k},j}\chi_j ,
\end{equation}
where the eigen vector $\chi_j$ is
\begin{eqnarray}
\chi_1 = \begin{bmatrix}
       \eta^1            \\[0.3em]
       \eta^1
     \end{bmatrix}\text{, } \chi_2 = \begin{bmatrix}
       \eta^2            \\[0.3em]
       -\eta^2
     \end{bmatrix}\text{, }\eta^1 =  \begin{bmatrix}
       1           \\[0.3em]
       0
     \end{bmatrix} \text{, } \eta^2=  \begin{bmatrix}
      0   \\[0.3em]
       1
     \end{bmatrix}
\end{eqnarray}

These spinors are the eigenvectors of $\gamma^0\gamma^3$ in the representation of $\gamma$ matrices. They are not exactly the same as the spin up and spin down eigen vectors. Since two eiegn vectors with positive eigenvalues are enough to cover the full space, we neglect the eigen vectors with negative eigen values\cite{Kluger:1992gb}. $\chi_s$s satisfy the relation
\begin{eqnarray}
\sum^4_{\alpha=1}(\chi^{\dagger}_r)^\alpha (\chi_s)_\alpha=2\delta_{rs} .
\end{eqnarray}

 The mode function $f_{\boldsymbol{k},j}$ satisfies
\begin{eqnarray}
\label{f-mode}
\frac{d^2 f_{\boldsymbol{k},j}}{dt^2}+\Big( \omega_{\boldsymbol{k}}^2-i e\frac{da}{dt}\Big) f_{\boldsymbol{k},j}=0 .
\end{eqnarray}

Here, $\omega_{\boldsymbol{k}}^2=p_3^2+\boldsymbol{k}_-^2+m^2$, $\boldsymbol{k}_-^2=k_1^2+k_2^2$ and $p_i=k^i-eA^i$. Eq. (\ref{f-mode}) is a second order differential equation, so each $j$ has two independent solutions, $f_{\boldsymbol{k},j}^+$ and $f_{\boldsymbol{k},j}^-$. Since Dirac equation has only 4 independent solutions, both $j=1,2$ sets can span a linearly independent set of Dirac solutions.  From  eq. (\ref{transform}) we have

\begin{equation}
\label{psitophi}
\psi^{\pm}_{\boldsymbol{k},j}=(i\gamma^0\partial_0+\gamma^i k_i-e\gamma^3 A_3+m)\phi_{\boldsymbol{k},j}^{\pm}
\end{equation}

 here, $\phi_{\boldsymbol{k},j}^{\pm}=e^{i \boldsymbol{k}\cdot \boldsymbol{x}}f_{\boldsymbol{k},j}^{\pm}\chi_j$.
After normalization, $\psi^{\pm}_{\boldsymbol{k},j}$ satisfy the relation \cite{Kluger:1992gb},

\begin{equation}
\psi^\pm_r{}^\dagger \psi^\pm_s =\delta_{rs}\text{, } \psi^\pm_r{}^\dagger \psi^\mp_s =0.
\end{equation}

The four $\psi^\pm_r{}$s are orthogonal to each other. The $\psi$ field is then quantized and written in the  form

\begin{eqnarray}
\label{field1}
\psi &=&\int \sum_{j=1,2}\Big[b_j (\boldsymbol{k})\psi_{\boldsymbol{k},j}^++d_j^\dagger(-\boldsymbol{k})\psi^-_{\boldsymbol{k}, j}\Big]\frac{d\boldsymbol{k}}{(2\pi)^3}
\end{eqnarray}
where $b_j$ and $d_j^\dagger$ are the electron annihilation and positron creation operator respectively. The operators satisfy the usual anti-commutation relation.
\begin{eqnarray}
\{ b_r(\boldsymbol{k}),b_s^\dagger(\boldsymbol{q}) \}=\{ d_r(\boldsymbol{k}),d_s^\dagger(\boldsymbol{q}) \}=(2\pi)^3\delta^3(\boldsymbol{k}-\boldsymbol{q})\delta_{rs}
\end{eqnarray}

Then the $\psi$ field also satisfies the anti-commutation relation,
\begin{equation}
\{ \psi_\alpha(t,\boldsymbol{x}),\psi_\beta^\dagger(t,\boldsymbol{y})) \}=\delta^3(\boldsymbol{x}-\boldsymbol{y})\delta_{\alpha \beta}.
\end{equation}

\section{ ``In" vacuum and ``Out" vacuum.}
The former section was about the second quantization of the $\psi$ field. However, this representation cannot give the spin of the particle pairs directly. We have to change the representation to up and down spinor bases,

\begin{eqnarray}
\label{field2}
\psi&=&\sum_{r=1,2}\int \Big[ b_r^{(0)}(\boldsymbol{k},t)u_{r,\boldsymbol{k}}e^{-i\int\omega_{\boldsymbol{k}} dt}\nonumber\\
&&+d_r^{(0)\dagger}(\boldsymbol{-k},t)v_{r,-\boldsymbol{k}}e^{i\int\omega_{\boldsymbol{k}} dt}\Big]e^{i\boldsymbol{k}\cdot \boldsymbol{x}}\frac{d\boldsymbol{k}}{(2\pi)^3}.
\end{eqnarray}
Here, $u_{r,\boldsymbol{k}}$ and $v_{r,-\boldsymbol{k}}$ are defined as
\begin{eqnarray}
\label{updownspinor}
u_{r,\boldsymbol{k}} =  \begin{bmatrix}
       \sqrt{\frac{\omega_{\boldsymbol{k}}+m}{2\omega_{\boldsymbol{k}}}}\eta^r           \\[0.3em]
       \frac{\vec{\sigma}\cdot \vec{p}}{\sqrt{2\omega_{\boldsymbol{k}}(\omega_{\boldsymbol{k}}+m)}}\eta^r
     \end{bmatrix} \text{, } v_{r,-\boldsymbol{k}} =  \begin{bmatrix}
      \frac{-\vec{\sigma}\cdot \vec{p}}{\sqrt{2\omega_{\boldsymbol{k}}(\omega_{\boldsymbol{k}}+m)}}\eta^r       \\[0.3em]
       \sqrt{\frac{\omega_{\boldsymbol{k}}+m}{2\omega_{\boldsymbol{k}}}}\eta^r
     \end{bmatrix}
\end{eqnarray}

where $u_{1,\boldsymbol{k}}$ and $u_{2,\boldsymbol{k}}$ are spin up and spin down electron spinor respectively (along z-direction). $\vec{p}=(k^1-eA^1,k^2-eA^2,k^3-eA^3 )$. Since we discuss only electric field in z direction, $A^1=A^2=0$. $v_{1,\boldsymbol{-k}}$ and $v_{2,-\boldsymbol{k}}$ are spin down and spin up positron spinors respectively.  $u_{r,\boldsymbol{k}}$ and $v_{r,\boldsymbol{k}}$ satisfy
\begin{equation}
u_{r,\boldsymbol{k}}^\dagger u_{r',\boldsymbol{k}}=\delta_{r,r'}\text{, } v_{r,\boldsymbol{k}}^\dagger v_{r',\boldsymbol{k}}=\delta_{r,r'}\text{, } u_{r,\boldsymbol{k}}^\dagger v_{r',-\boldsymbol{k}}=0 .
\end{equation}

One can relate eq (\ref{field1}) and eq. (\ref{field2}) with the Bogoliubov transformation

\begin{eqnarray}
\label{b1}
b_r^{(0)}(\boldsymbol{k},t)&=&\sum_{s=1,2}\alpha^s_{\boldsymbol{k},r}(t)b_s(\boldsymbol{k})+\beta^s_{\boldsymbol{k},r}(t)d_s(-\boldsymbol{k})^\dagger\\
\label{d1}
d_r^{(0)}(-\boldsymbol{k},t)^\dagger&=&\sum_{s=1,2}-\beta^{*s}_{\boldsymbol{k},r}(t)b_s(\boldsymbol{k})+\alpha^{*s}_{\boldsymbol{k},r}(t)d_s(-\boldsymbol{k})^\dagger
\end{eqnarray}

From the canonical anti-communication relation, one finds

\begin{equation}
\sum_{r=1,2}(|\alpha^s_{\boldsymbol{k},r}|^2+|\beta^s_{\boldsymbol{k},r}|^2)=1
\end{equation}

Once Bogoliubov transformation is substituted in eq. (\ref{field2}), $\psi^+_{\boldsymbol{k},s}$ and $\psi^-_{\boldsymbol{k},s}$ are found by comparing this equation with eq. (\ref{field1}),

\begin{eqnarray}
\label{factor1}
\psi^+_{\boldsymbol{k},s}&=&\sum_{r=1,2}\alpha^s_{\boldsymbol{k},r}u_{r,\boldsymbol{k}}e^{-i\int \omega_{\boldsymbol{k}}dt}-\beta^{*s}_{\boldsymbol{k},r}v_{r,-\boldsymbol{k}}e^{i\int \omega_{\boldsymbol{k}}dt}\\
\label{factor2}
\psi^-_{\boldsymbol{k},s}&=&\sum_{r=1,2}\beta^s_{\boldsymbol{k},r}u_{r,\boldsymbol{k}}e^{-i\int \omega_{\boldsymbol{k}}dt}+\alpha^{*s}_{\boldsymbol{k},r}v_{r,-\boldsymbol{k}}e^{i\int \omega_{\boldsymbol{k}}dt}
\end{eqnarray}

As usual, the number of particles produced per unit phase space volume at a given momentum is given by

\begin{eqnarray}
n(\boldsymbol{k},t)&=&\sum_{r=1,2}\bra{0,\text{in}}b_r^{(0)\dagger}(\boldsymbol{k},t)b_r^{(0)}(\boldsymbol{k},t)\ket{0,\text{in}}\nonumber\\
\label{number}
&=&\sum_{s=1,2;r=1,2}|\beta_{\boldsymbol{k},r}^s(t)|^2
\end{eqnarray}




From eq. (\ref{factor1}), $\alpha_{\boldsymbol{k},r}^{s}$ and $\beta_{\boldsymbol{k},r}^{s}$ can be found in terms of $\psi^+_{\boldsymbol{k},s}$.

\begin{eqnarray}
-\beta_{\boldsymbol{k},r}^{*s}e^{i\int \omega_{\boldsymbol{k}}dt}&=& v_{r,-\boldsymbol{k}}^\dagger \psi^+_{\boldsymbol{k},s}\\
\alpha_{\boldsymbol{k},r}^{s}e^{-i\int \omega_{\boldsymbol{k}}dt}&=& u_{r,\boldsymbol{k}}^\dagger \psi^+_{\boldsymbol{k},s}
\end{eqnarray}

One can substitute eq. (\ref{psitophi}) and (\ref{updownspinor}) to obtain $\alpha_{\boldsymbol{k},r}^{s}$ and $\beta_{\boldsymbol{k},r}^{s}$,

\begin{eqnarray}
\beta_{\boldsymbol{k},1}^{*1}&=& -e^{-i\int \omega_{\boldsymbol{k}}dt}\frac{(\omega_{\boldsymbol{k}}+m+p_3)(\omega_{\boldsymbol{k}}f^+_{\boldsymbol{k},1}-i\dot{f}^+_{\boldsymbol{k},1})}{\sqrt{2\omega_{\boldsymbol{k}}(\omega_{\boldsymbol{k}}+m)}}\\
\beta_{\boldsymbol{k},2}^{*1}&=&-e^{-i\int \omega_{\boldsymbol{k}}dt} \frac{(p_1+ip_2)(\omega_{\boldsymbol{k}}f^+_{\boldsymbol{k},1}-i\dot{f}^+_{\boldsymbol{k},1})}{\sqrt{2\omega_{\boldsymbol{k}}(\omega_{\boldsymbol{k}}+m)}}\\
\alpha_{\boldsymbol{k},1}^{1}&=& e^{i\int \omega_{\boldsymbol{k}}dt}\frac{(\omega_{\boldsymbol{k}}+m-p_3)(\omega_{\boldsymbol{k}}f^+_{\boldsymbol{k},1}+i\dot{f}^+_{\boldsymbol{k},1})}{\sqrt{2\omega_{\boldsymbol{k}}(\omega_{\boldsymbol{k}}+m)}}\\
\alpha_{\boldsymbol{k},2}^{1}&=& e^{i\int \omega_{\boldsymbol{k}}dt}\frac{-(p_1+ip_2)(\omega_{\boldsymbol{k}}f^+_{\boldsymbol{k},1}+i\dot{f}^+_{\boldsymbol{k},1})}{\sqrt{2\omega_{\boldsymbol{k}}(\omega_{\boldsymbol{k}}+m)}}\\
\beta_{\boldsymbol{k},1}^{*2}&=&-e^{-i\int \omega_{\boldsymbol{k}}dt} \frac{(p_1-ip_2)(\omega_{\boldsymbol{k}}f^+_{\boldsymbol{k},2}-i\dot{f}^+_{\boldsymbol{k},2})}{\sqrt{2\omega_{\boldsymbol{k}}(\omega_{\boldsymbol{k}}+m)}}\\
\beta_{\boldsymbol{k},2}^{*2}&=& e^{-i\int \omega_{\boldsymbol{k}}dt}\frac{(\omega_{\boldsymbol{k}}+m+p_3)(\omega_{\boldsymbol{k}}f^+_{\boldsymbol{k},2}-i\dot{f}^+_{\boldsymbol{k},2})}{\sqrt{2\omega_{\boldsymbol{k}}(\omega_{\boldsymbol{k}}+m)}}\\
\alpha_{\boldsymbol{k},1}^{2}&=& e^{i\int \omega_{\boldsymbol{k}}dt}\frac{(p_1-ip_2)(\omega_{\boldsymbol{k}}f^+_{\boldsymbol{k},2}+i\dot{f}^+_{\boldsymbol{k},2})}{\sqrt{2\omega_{\boldsymbol{k}}(\omega_{\boldsymbol{k}}+m)}}\\
\alpha_{\boldsymbol{k},2}^{2}&=&e^{i\int \omega_{\boldsymbol{k}}dt}\frac{(\omega_{\boldsymbol{k}}+m-p_3)(\omega_{\boldsymbol{k}}f^+_{\boldsymbol{k},2}+i\dot{f}^+_{\boldsymbol{k},2})}{\sqrt{2\omega_{\boldsymbol{k}}(\omega_{\boldsymbol{k}}+m)}}
\end{eqnarray}

As $t\rightarrow -\infty$, $f^+_{\boldsymbol{k},s}\propto e^{-i\omega_{\boldsymbol{k}} t}$ and $\omega_{\boldsymbol{k}}f^+_{\boldsymbol{k},r}-i\dot{f}^+_{\boldsymbol{k},r}=0$. All $\beta_{\boldsymbol{k},s}$ are $0$. According to eq. (\ref{number}), $n=0$ and there is no particle creation. 
Since $\alpha_{\boldsymbol{k},i}^{j}$s in general are not both $0$, the ``out" vacuum's spinors are not the same as the ``in" vacuum's spinors.
Under the Bogoliubov transform the ``in" vacuum, $|\text{in}>$,  is annihilated by  $b_s(\boldsymbol{k})$ and $d_s(-\boldsymbol{k})$

 \begin{eqnarray}
b_s(\boldsymbol{k})\ket{0,\text{in}}=d_s(-\boldsymbol{k})\ket{0,\text{in}}=0
\end{eqnarray}
while the ``out" vacuum, $\ket{0,\text{out}}$, is annihilated by $b_r^{(0)}(\boldsymbol{k},t)$ and $d_r^{(0)}(-\boldsymbol{k},t)$,

 \begin{eqnarray}
 \label{out1}
b_r^{(0)}(\boldsymbol{k},t)\ket{0,\text{out}}=d_r^{(0)}(-\boldsymbol{k},t)\ket{0,\text{out}}=0
\end{eqnarray}

We may write the Bogoliubov transform as a unitary transform, $U_{\boldsymbol{k}}$,

 \begin{eqnarray}
b_r^{(0)}(\boldsymbol{k},t)&=&U_{\boldsymbol{k}} b_r^{(0)}(\boldsymbol{k},-\infty)U_{\boldsymbol{k}}^\dagger\\
d_r^{(0)}(-\boldsymbol{k},t)^\dagger&=&U_{\boldsymbol{k}} d_r^{(0)}(-\boldsymbol{k},-\infty)^\dagger U_{\boldsymbol{k}}^\dagger
\end{eqnarray}

$U_{\boldsymbol{k}}$ can relate the ``out" vacuum to the ``in" vacuum as

\begin{eqnarray}
\ket{0,\text{out}}=U_{\boldsymbol{k}}\ket{0,\text{in}}
\end{eqnarray}

This is a 4 mode transform. Its complete form is complicated\cite{1990PhRvA..41.4625M,2001quant.ph..9020Q}, but we do not need the complete transform. We only need to write the ``out" vacuum in terms of $b_i(\boldsymbol{k})$, $d_j(\boldsymbol{k})$ operators, and the ``in" vacuum.  The ``out" vacuum can be written in the form
\begin{equation}
\ket{0,\text{out}}=\prod_{\boldsymbol{k},s} A\exp(\sum_{ij}B_{ij}b_i^\dagger d_j^\dagger )\ket{0,\text{in}}
\end{equation}
$A$ and $B_{ij}$ can be found from eq. (\ref{out1}) and $\bra{0,\text{out}}\ket{0,\text{out}}=1$,

\begin{eqnarray}
B_{ij}&=&(-1)^{m}\frac{\alpha^m_{\boldsymbol{k},2}(t)\beta^j_{\boldsymbol{k},1}(t)- \alpha^m_{\boldsymbol{k},1}(t)\beta^j_{\boldsymbol{k},2}(t)}{\alpha^2_{\boldsymbol{k},2}(t)\alpha^1_{\boldsymbol{k},1}(t)- \alpha^2_{\boldsymbol{k},1}(t)\alpha^1_{\boldsymbol{k},2}(t)}\\
A&=&\sqrt{1+(|B_{11}|+|B_{12}|+|B_{21}|+|B_{22}|)^2} .
\end{eqnarray}

Here $m=1$, if $i=2$ and $m=2$ if $i=1$. One must be careful that $U_{\boldsymbol{k}}\neq \prod_{\boldsymbol{k},s} A\exp(\sum_{ij}B_{ij}b_i^\dagger d_j^\dagger )$, since some of $U_{\boldsymbol{k}}$ operators disappear while they operate on $\ket{0,\text{in}}$.

\section{ Observation of the spin up and spin down states.}
We are looking for the creation of a two particle state in the ``out" vacuum, $b^{(0)}_r (\boldsymbol{k},t)^\dagger d^{(-\boldsymbol{k},0)}_s(t)^\dagger \ket{\text{out}}$. This amplitude is
\begin{eqnarray}
T_{rs}&=&\bra{\text{out}} b^{(0)}_r (\boldsymbol{k},t) d^{(0)}_s(-\boldsymbol{k},t) \ket{\text{in}}\nonumber\\
&=&A\sum_{i=1,2} \beta^{i}_{\boldsymbol{k},r}(t)\Big(\alpha^{i}_{\boldsymbol{k},s}(t) -\sum_{j=1,2}\beta^{j}_{\boldsymbol{k},s}(t)B_{ij}^*\Big)
\end{eqnarray}
We can find the value of each component
\begin{eqnarray}
T_{11}&=&-\frac{\omega_{\boldsymbol{k}}^2+m\omega_{\boldsymbol{k}}-p_3^2}{\omega_{\boldsymbol{k}}(\omega_{\boldsymbol{k}}+m)}E +O(\beta^3)\\
T_{12}&=&\frac{p_3(p_1+ip_2)}{\omega_{\boldsymbol{k}}(\omega_{\boldsymbol{k}}+m)}E+O(\beta^3)\\
T_{21}&=&\frac{p_3(p_1-ip_2)}{\omega_{\boldsymbol{k}}(\omega_{\boldsymbol{k}}+m)}E+O(\beta^3)\\
T_{22}&=&\frac{\omega_{\boldsymbol{k}}^2+m\omega_{\boldsymbol{k}}-p_3^2}{\omega_{\boldsymbol{k}}(\omega_{\boldsymbol{k}}+m)}E+O(\beta^3)\\
E&=&(\omega_{\boldsymbol{k}}\bar{f}^+_{\boldsymbol{k},1}+i\dot{\bar{f}}^+_{\boldsymbol{k},1})(\omega_{\boldsymbol{k}}f^+_{\boldsymbol{k},1}+i\dot{f}^+_{\boldsymbol{k},1})e^{2i\int \omega_{\boldsymbol{k}} dt}A .
\end{eqnarray}

Here $\beta \propto \omega_{\boldsymbol{k}}f^{+*}_{\boldsymbol{k},1}+i\dot{f}^{+*}_{\boldsymbol{k},1}$ and $f^+_{\boldsymbol{k},1}=f^+_{\boldsymbol{k},2}$ is applied. If the particle creation is not very strong, then $O(\beta^3)$ is much smaller than $E$ which is proportional to $\beta$, so it can be neglected. We do not write the precise form down because it is complicated.  If the first index is $1(2)$, it creates an electron with spin up(down). If the second index is $1(2)$, it creates a positron with spin down(up). If the transverse momenta are $0$ ($p_1=0$ and $p_2=0$), only $T_{11}$ and $T_{22}$ are nonzero, which means the electron and positron have opposite spin orientations.  It means that  $A_2$ and $A_3$ in eq. (\ref{portion}) are both $0$, and the state is a maximally entangled Bell state (as in eq.(\ref{entangle})).    However, if $p_1$ or $p_2$ are not $0$, then $T_{12}$ and $T_{21}$ are not zero, and the electron and positron can have the same spin orientation.  All $A_i$ in eq. \eqref{portion} are not $0$ and the state is not a maximally entangled Bell state.  In other words, if one knows one of the particle's state, he still cannot determine the state of the other particle if the transverse momentum is not $0$. This proves that members of a Schwinger pair are not always entangled.

 We can also study the same effect in the propagation direction. The helicity eigen spinor can be written as a combination of spin up and spin down spinors and the creation and annihilation operator can be found by comparing the $\psi$ field component in this two bases.
The amplitudes for each case are


\begin{eqnarray}
T_{RR}&\approx&\frac{\cos\theta(\omega_{\boldsymbol{k}}^2+m\omega_{\boldsymbol{k}}-p_3^2)-p_3 p\sin^2\theta}{\omega_{\boldsymbol{k}}(\omega_{\boldsymbol{k}}+m)}E\\
T_{RL}&\approx&\frac{-\sin\theta(\omega_{\boldsymbol{k}}^2+m\omega_{\boldsymbol{k}}-p_3^2)-p_3 p\sin\theta\cos\theta}{\omega_{\boldsymbol{k}}(\omega_{\boldsymbol{k}}+m)}E\\
T_{LR}&\approx&\frac{-\sin\theta(\omega_{\boldsymbol{k}}^2+m\omega_{\boldsymbol{k}}-p_3^2)-p_3 p\sin\theta\cos\theta}{\omega_{\boldsymbol{k}}(\omega_{\boldsymbol{k}}+m)}E\\
T_{LL}&\approx&\frac{-\cos\theta(\omega_{\boldsymbol{k}}^2+m\omega_{\boldsymbol{k}}-p_3^2)+p_3 p\sin^2\theta}{\omega_{\boldsymbol{k}}(\omega_{\boldsymbol{k}}+m)}E
\end{eqnarray}

Here, $\cos\theta=p_3/p$ and $p^2=p_1^2+p_2^2+p_3^2$.  $R$($L$) is the right(left) handed helicity.

Again we neglected the terms proportional to $\beta^3$. $T_{RL}=T_{LR}=0$, if $p_1=p_2=0$. This pair is entangled. However, if there is transverse momentum ($p_1\neq 0$ or $p_2\neq 0$), then $T_{RL}$ and $T_{LR}$ are not $0$. If one knows one particle's helicity, he still cannot know the other particle's helicity. This proves that helicities od a Schwinger pair are not completely entangled.

\section{Conclusion}
We showed that a Schwinger particle pair's spins are not necessary in a maximally entangled Bell state. If the particle pair's momenta are not parallel to the electric field, then one cannot know the particle pair's spin states just by measuring the properties of one particle. In addition, if the external field is spatially non-uniform, the particle pair can also gain linear momentum\cite{2008PhRvD..78j5013K}. Then both linear and angular momenta of the pair are not completely correlated. To make a precise statement, members of a virtual particle pair generated by vacuum fluctuations are highly correlated. However, for these particles to become real, they have to interact with an external field. This interaction ruins the original correlation.

 The Schwinger pair production is expected to occur when the electric field is above the Schwinger limit, $E_s=1.32\times 10^{18}V/m$. This is far beyond the current laser system's limit ($\sim 10^{13}$ to $10^{14} V/m$)\cite{2006RvMP...78..309M,2008OExpr..16.2109Y}. However, recent development of ultrashort and ultraintense laser pulse raise the possibility to approach the threshold in the foreseeable future\cite{2002PhRvS...5c1301T}. Especially, nonlinear QED effect as $e^+e^-$ pair photoproduction by hard photon\cite{1997PhRvL..79.1626B,Bell:2008hk,2011PhRvL.106c5001N} and nonlinear Compton scattering have been observed at laser intensity $I=10^{22}W/m^2$\cite{1996PhRvL..76.3116B}. It has been shown that multiple colliding electromagnetic pulses can even lower the laser's intensity threshold of $e^+e^-$ pair production to $10^{26}W/cm^2$\cite{Dunne:2009gi,Bulanov:2010ei}, which is much lower than the Schwinger pair production threshold, $10^{29}W/cm^2$.  There are several projects to achieve intensity $10^{26}$ to $10^{28}W/cm^2$\cite{2006RvMP...78..309M,2006NatPh...2....2D,Korn:2009}. $e^+e^-$ pair production by multiple laser pulses has been proposed in the new lasers systems, such as Extreme Light Infrastructure (ELI)\cite{Korn:2009} and  the European High Power laser Energy Research facility (HiPER)\cite{2006NatPh...2....2D}. This makes observation of the Schwinger pairs quite possible. The pair production is tightly related to the focused laser's geometric structure and polarization, and not only on the energy input\cite{Banerjee:2018gyt}. In our case we study the particle pair's spin correlation. We will find that the spin correlation depends on the scattering angle. This directly implies that the pair's state is not just one of the Bell entangled states, which is sharply different from what is generally expected.    

Hawking radiation is made of particle pairs generated by vacuum fluctuations. The negative energy particle falls into the horizon, while the positive energy one leaves the horizon and is radiated away. This particle pair is assumed to be entangled based on the locality assumption\cite{Mathur:2009hf}. However, the particle pair is generated by a similar process as a Schwinger pair production, which implies that Hawking radiation particle pair is not completely correlated. This is true in general, since the process of Hawking radiation takes away angular momentum from a black hole, and the products of radiation do not move in the radial direction.
In addition, the wavelength of emitted particles is about the radius of the black hole. Thus, the external gravitational field cannot be described by a uniform distribution. The produced particles can gain linear momenta from external field, and according to our study, their spins are not completely correlated. This implies the semiclassical gravity (and perhaps full quantum gravity) is not a completely local phenomenon.

\begin{acknowledgments}
D.C Dai was supported by the National Science Foundation of China (Grant No. 11433001 and 11775140) and National Basic Research Program of China (973 Program 2015CB857001).

\end{acknowledgments}

\end{document}